\documentstyle[prl,aps,epsf,multicol]{revtex}
\begin{document}
\draft
 
\title{Quasiparticle Localization Transition in Dirty Superconductors}

\author{Smitha Vishveshwara$^1$ and Matthew P. A. Fisher$^2$}
\address{$^1$Department of Physics, University of California, Santa Barbara, CA 93106 \\
$^2$Institute for Theoretical Physics, University of California,
Santa Barbara, CA 93106--4030
}

\date{\today}
\maketitle

\begin{abstract}
In this paper, a delocalization-localization transition within the superconducting state is explored. The symmetries of the Bogoliubov deGennes Hamiltonian endow the two associated superconducting phases - the thermal metal and the thermal insulator-, and the critical point between them  with properties that are in stark contrast with analogous phases in normal systems. Here, for systems preserving spin rotational invariance and time reversal symmetry in three dimensions, the transition is shown to be of a different universality class from its normal partner  by extracting and comparing the localization length exponent. The density of states, which may be regarded as the 'order parameter' for the field theory describing the superconducting system, is studied for its unusual properties about criticality.

\end{abstract}

\vspace{0.15cm}


\begin{multicols}{2}
\narrowtext 

\section{Introduction}
The problem of Anderson localization in normal disordered electronic systems has been studied for years, and has continued to pose intriguing puzzles\cite{PLTVR,Bodo} . Recently, the question of whether an analogous transition can occur in the behaviour of low-energy quasiparticle excitations about a superconducting ground state has come to light (see for e.g. Ref.\cite{Bund}.) In fact, the stability of two very distinct superconducting phases which may be characterized by their transport properties - one with extended quasiparticle states at the Fermi energy, and the other with localized states- has been established\cite{TSMF1}. Here, it is the transition point between these two phases, and its critical properties, that forms the subject of our attention.

While one can afford to ask of the superconducting systems the same questions that have been addressed in normal systems, the former displays refreshingly new physics that shows both conceptual and qualitative differences from the latter. To begin with, quasiparticle excitations in superconducting systems do not conserve charge, and thus one cannot study the transition through charge dynamics. This absence of $U(1)$ symmetry in the Bogoliubov deGennes Hamiltonian, which we employ to describe the excitations in question, has marked consequences. As emphasized by Ref.\cite{AZ}, it gives rise to a host of new universality classes ,with critical exponents whose values  are significantly different from those of their normal partners. In particular, the density of states(DOS) exhibits astonishing features. In this paper, we explore these surprises in the context of superconducting systems that respect spin-rotational ($SU(2)$) invariance and time-reversal (T) symmetry.

The issue of a delocalization-localization transition within the superconducting state has been a recurrent theme (see for e.g. Ref\cite{Opp}). But it is only in recent years that analyses of both microscopic and continuum models have paid focused attention from a variety of different avenues\cite{Bund,TSMF1,Ilya,TSMF2,TSMF3,SVSF,TSMF4}, and delved into the prospect of making this transition realizable in physical systems. To repeat the example offered of the dirty d-wave superconductor, consider a system of impure superconducting sheets with d-wave pairing coupled to one another. At the nodal points, one has low-energy quasiparticle excitations, and in fact, due to disorder, one even expects a finite DOS at the Fermi energy\cite{Gork}. For low interplane coupling strength or high impurity concentration, one would expect these states to be localized, and upon increasing the coupling or lowering the disorder, one could conceive of accessing a critical point beyond which these states become extended. Here, we study such a transition on more generic grounds by exploiting the field-theoretic set-up offered by Ref.\cite{TSMF1},  and through numerical analyses, both of which serve to bring out the novel features of superconducting systems quite dramatically.

We review the framework used to describe quasiparticle excitations  in the absence of interactions. We expand on some details of the two phases, the 'thermal metal' and the 'thermal insulator'. Then, in order to keep our paper self-contained, we elaborate on the field-theoretic and numerical methods. We proceed to discuss the field-theoretic and numerical predictions for the localization length exponent $\nu$ associated with the thermal metal-thermal insulator transition, and show that they both confirm the existence of a new universality class with $\nu<\nu_n$, where $\nu_n$ is the corresponding exponent for normal systems. We then study the unique properties of the DOS at criticality. Finally, we mention the characteristics of transitions in superconducting systems besides those that respect $SU(2)$ and T symmetry, and possibilities for experiment.

\section{Models and Symmetries}

\subsection{The Superconducting Hamiltonian}
Quasiparticle excitations about the superconducting ground state are well described within the framework of the Bogoliubov deGennes(BdG) Hamiltonian.  For the spin-singlet paired superconductor, which we focus on here, the BdG Hamiltonian has the form
\begin{eqnarray}
&{\cal H}_0  =  H_1 + H_2  ,&\nonumber \\
H_1 & & \nonumber \\
= & \int d^d x c^{\dagger}_{\sigma}(x)\left(\frac{\left(-i\hbar\vec \nabla -\frac{e}{c}\vec A(x)\right)^2}{2m}
 - E_F + V(x) \right) c_{\sigma}(x), & \nonumber \\ 
H_2 & & \nonumber \\= & \int d^dx d^dx' (c^\dagger_{\uparrow} (x)\Delta(x,x')c^{\dagger}_{\downarrow} (x')  +  
c_{\downarrow} (x)\Delta^*(x,x') c_{\uparrow} (x')),& \nonumber
\end{eqnarray}
where $c^{\dagger}$ and $c$ are electron creation and annhilation operators respectively, $m$ the mass , $E_F$ the Fermi Energy , $V(x)$ a random potential describing impurities in the system, and $A(x)$ a vector potential describing any external magnetic field. The lattice version of ${\cal H}_0$, which is more tractable for numerics, has the form
\begin{equation}
{\cal H}_{0L} = \sum_{ij}[t_{ij}\sum_{\alpha}c^{\dagger}_{i\alpha}c_{j\alpha}
+ \left(\Delta_{ij}c^{\dagger}_{i\uparrow}c^{\dagger}_{j\downarrow} + h.c \right)] .
\label{HL} 
\end{equation}
From Hermiticity, one requires the condition $t_{ij} = t_{ji}^*$, while spin rotation invariance requires $\Delta_{ij} = \Delta_{ji}$. The BdG Hamiltonian contains anamolous terms that reflect the fact that the excitations do not conserve charge, and hence break the associated $U(1)$ symmetry. As a result, the BdG Hamiltonian lives in an 'extended particle-hole space' with twice the degrees of freedom of normal electronic systems. More importantly, the absence of $U(1)$ charge conservation endows the BdG Hamiltonian with symmetry properties which are completely different from normal systems. It is this difference that plays the key role in giving rise to properties that are unique to superconducting systems. 

The symmetries of the BdG Hamiltonian have recently been studied in the context of mesoscopic systems and random matrix theory\cite{AZ}. Within the class of models described in Ref\cite{AZ}, the Hamiltonians that we study in this paper have SU(2) and T symmetry; they describe systems that are singlet-paired and have no spin-orbit scattering, thus having spin-rotational invariance, and in addition, they have time-reversal invariance.

Since the quasiparticles conserve both spin and energy, one can explicitly write Eq.\ref{HL} in terms of conserved quantities by defining a new set of fermionic $d-$operators:
\begin{equation}
\label{c_to_d}
d_{i\uparrow} = c_{i\uparrow},~~ d_{i\downarrow} = c^{\dagger}_{i\downarrow } .
\end{equation}
Thus one makes a particle-hole transformation on the spin-down operators, but leaves the spin-up operators unchanged.
The Hamiltonian of Eq.\ref{HL} now takes the form
\begin{equation}
\label{HdL}
{\cal H}_L = \sum_{ij} d^{\dagger}_i \left(\begin{array}{cc}t_{ij} & \Delta_{ij}
\\
                                                \Delta_{ij}^* & -t_{ij}^*
                                                               \end{array}
\right) d_j
                                     \equiv \sum_{ij}d_{i}^{\dagger}H_{ij}d_j ,
\end{equation} 
 where time reversal invariance requires $t_{ij} $ and $\Delta_{ij}$ be real. 
 Spin rotational invariance now requires 

\begin{equation}
\sigma^y H_{ij} \sigma^y = -H_{ij}^*,
\label{spinrot}
\end{equation} 
where $\sigma^y$ is the standard Pauli matrix. Spin conservation along the z-direction is evident from the fact that the physical spin is related to the number operator for the 'd-particles' by
\begin{equation}
\label{spin}
S^z_i = \frac{\hbar}{2}\left(d^{\dagger}_id_i -1 \right) ,
\end{equation}
and that ${\cal H}_L$ of Eq.\ref{HdL} conserves particle number. 

 The Hamiltonian ${\cal H}_L$, in principle, may be diagonalized by solving the eigenvalue equation

\begin{equation}
\label{BdiaG}
\sum_i H_{ij} \left[\begin{array}{c} u(j) \\
                                     v(j) \end{array}\right]
                         = E  \left[\begin{array}{c} u(i) \\
                                     v(i) \end{array}\right] ,
\label{diag}
\end{equation}
where $u$ and $v$ describe the wave-function amplitudes at each site. Given Eq.\ref{diag}, one can construct the following state

\begin{equation}
 i \sigma^y \left[\begin{array}{c} u(j)^* \\
                                     v(j)^* \end{array}\right]
                         =  \left[\begin{array}{c} v^*(i) \\
                                     -u^*(i) \end{array}\right],
\label{srstate}
\end{equation}
with eigenvalue $-E$; $SU(2)$ invariance requires a symmetric dispersion about the Fermi energy, and eigenvalues of the BdG Hamiltonian come in pairs $(E,-E)$.

 We now draw attention to the physical situation described by the wavefunctions of Eq.\ref{diag}, and the relation they bear to the phase transition of interest. To begin with, in gapless superconductors, which we focus on here, one finds states at and about the Fermi energy. These states have a profound impact on thermodynamic and transport properties, as discussed previously\cite{TSMF1,TSMF2,SVSF}. Moreover, the transport properties depend crucially on the nature of the eigenstates of Eq.\ref{diag} at $E=0$, which in turn is determined by the spatial configuration of the random potential $V(x)$ and the gap-function $\Delta(x)$. In particular, depending on whether these eigenstates are extended or  localized, one can conceive of two very different superconducting phases - the 'thermal metal' which is capable of transporting energy, and the 'thermal insulator' which cannot conduct energy over large length scales. In fact, it has been shown that both phases are stable in three dimensions\cite{TSMF1}. Thus, one can characterize the two phases by the thermal conductivity $\kappa$. In systems with SU(2) symmetry, since the same quasiparticles that carry energy also carry spin, the spin conductivity $\sigma_s$, may also be used to describe the two phases, and is related to $\kappa$ via an analog of the Weidemann-Franz law in the thermal metal:
\begin{equation}
\frac{\kappa}{T\sigma_s} = const.
\label{WF}
\end{equation}
Note however, that as the quasiparticles do not carry well defined charge, the Weidemann-Franz law breaks down with regards to electrical conductivity. 

The two distinct superconducting phases at hand are in analogy with, but quite different from, the metallic and insulating phases in normal systems. We study the critical properties of the transition between the two phases using the formalism and physical set-up described below.

\subsection{A Field-theoretic Formulation}
\label{FTF}
As with the case of Anderson localization\cite{Wegner,Bodo}, the problem of quasiparticle localization within the superconducting state can be described within a field-theoretic framework. To briefly review its key features,
in previous work\cite{TSMF1}, starting with the Hamiltonian of Eq.\ref{HL}, T.Senthil.et.al derived a useful field theoretic action from which one can extract a rich variety of properties of superconducting systems. Coupling the fermionic degrees of freedom in Eq.\ref{HL} to an infinitesimal Zeeman field $\eta$ (which acts as a chemical potential for the d-quasiparticles), and employing the replica method to calculate disorder averages, they obtained an effective action. Fluctuations about the saddle-point of this action were captured near two spatial dimensions by a non-linear sigma model($NL \sigma M$) treatment, yielding for the final form of the action,
\begin{equation}
S_{N L \sigma M} = \int d^d x [\frac{1}{2 t}Tr(\vec{\nabla U} .\vec{\nabla U^{\dagger}} ) - \eta Tr(U + U^{\dagger})].
\label{NLSM}
\end{equation}
Here, 't' is  a dimensionless coupling constant that has the physical interpretation of inverse spin conductance, i.e.  $\frac{1}{t} = \frac{\pi}{2} \sigma_s$. $U(x)$ is a matrix with  symplectic $Sp(2n)$ group structure, where 'n' is the number of replicas.

The action $S_{NL \sigma M}$ given above is quite different from the analogous action describing normal systems. The field theory , referred to as the 'the principle chiral $Sp(2n)$ model', has its first term invariant under the global 'rotation' $U \rightarrow A^{\dagger} U B$, where $A^{\dagger} , B \in Sp(2n)$ ,  thus possessesing $Sp(2n) \times Sp(2n)$ symmetry. The second term in the action reduces the symmetry to an $Sp(2n)$ symmetry as it only allows invariance under $U \rightarrow A^{\dagger} U A$. A knowledge of this symmetry structure proves to be very useful in deriving critical properties.

As done for electrical conductivity in normal systems, we employ a scaling theory for the inverse spin conductivity $t$, and analyze the critical point seperating the thermal metal and the thermal insulator. Specifically, we extract the localization length exponent and the unusual singular behaviour of the DOS at the Fermi energy.

\subsection{Hamiltonian for Numerics}
The predictive power of the effective field theory of the previous section lies in the fact that some results derived from it are universal to all Hamiltonians satisfying the appropriate symmetries. In practice, we find that in obtaining the localization length exponent, an analog of the tight-binding Anderson model commonly used for numerics\cite{KMcK,Tero} shows crisp data with much less noise than other models that we have studied. We focus on the same model for the density of states since the localisation length numerics enables us to identify the critical point quite accurately. With reference to Eq.\ref{HL}, the couplings take the form
\begin{equation}
 t_{ij} = \left\{ \begin{array}{r@{\quad , \quad}l} \frac{1}{\sqrt{2}} & i\ j \, ,n.n. \\ V_{it} & i = j \\ 0 & otherwise\end{array} \right .
\label{cupt}
\end{equation}
\begin{equation}
\Delta_{ij} = \left\{ \begin{array}{r@{\quad , \quad}l} \frac{1}{\sqrt{2}} & i\ j \, ,n.n. \\ V_{i \Delta} & i = j \\ 0 & otherwise\end{array} \right. 
\label{cupd}
\end{equation}
 where n.n denotes nearest neighbours, and $V_{it}$ and $V_{i \Delta}$ are on-site random variables chosen from a uniform probability distribution ranging from $-W$ to $+W$. We work with a three-dimensional cubic lattice described by Eq.\ref{HL}, and having real couplings with the specific form
\begin{equation}
{\cal H}_L =  \frac{1}{\sqrt{2}}[t \otimes \sigma^z + \Delta \otimes \sigma^x].
\label{Hmat}
\end{equation}
Here, the $\sigma$'s denote Pauli matrices, and $t$ and $\Delta$ are matrices with off-diagonal terms set to unity, and diagonal terms taking on values $\sqrt{2} V_{it}$ and $\sqrt{2} V_{i \Delta}$.
\subsection{The Transfer Matrix}
\label{TM}
While we obtain the DOS from the Hamiltonian of Eq.\ref{Hmat} by the straightforward process of exact diagonalization, we extract the localization length exponent by means of a transfer matrix formulation that caters to Eq.\ref{Hmat} and tremendously reduces the dimensions of the matrices involved in numerical work.

\begin{figure}
\epsfxsize=3.5in
\centerline{\epsffile{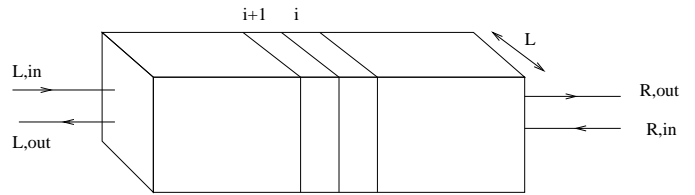}}
\vspace{0.15in}
\caption{A quasi 1-dimensional set-up for transfer matrices. In-going and out-going wavefunction amplitudes at either end are denoted by $\vec{\psi}_{in}$ and $\vec{\psi}_{out}$ respectively. }
\vspace{0.15in}
\label{quasi}
\end{figure}

To describe its principle, consider a quasi 1-dimensional strip in d-dimensions with cross-sectional area $L^{d-1}$, and in-going and out-going states at either end of this long strip as shown in Fig.\ref{quasi}. One might formally obtain the scattering matrix S for the in-going and out-going states using the definition
\begin{equation}
S = \lim_{T\to\infty}\exp{\frac{-iHT}{\hbar}}.
\label{scat1}
\end{equation}
The scattering matrix, in the specific basis of in-going and out-going states can be written in terms of reflection and transmission matrices 'r' and 't' respectively\cite{Been}:
\begin{equation}
S = \left(\begin{array}{cc}r & t'\\t & r'\end{array}\right).
\label{scat2}
\end{equation}
Given this form of the scattering matrix, one can easily derive the transfer matrix T which has the property
\begin{equation}
\vec{\psi_L} = T \vec{\psi_R}.
\end{equation}
The symmetries of the BdG Hamiltonian imply that the transfer matrix too has very specific symmetry properties which distinguish it from those of normal systems. Of late, these symmetries have been explored on group theoretic grounds\cite{Caselle}.

The transfer matrix T can be constructed by multiplying a set of transfer matrices, each appropriate for a slice of the strip shown in Fig.\ref{quasi}:
\begin{equation}
\vec{\psi}_{i+1,R} = T_i\vec{\psi}_{i,R},
\end{equation}

\begin{equation}
T = \prod_{i=1}^{N}T_i.
\label{Tmult}
\end{equation}

The form of the transfer matrix that we use for numerical calculation does not make the symmetry of the BdG Hamiltonian manifest, but it is tailored specifically for a tight-binding Hamiltonian such as the one described in Eq.\ref{HL}. We begin by writing the Schrodinger Eq.\ref{diag} as a difference equation
\begin{equation}
A_i\vec{D}_{i} + B_{i,i+1}\vec{D}_{i+1} + B_{i,i-1}\vec{D}_{i-1}  = E\vec{D}_{i},
\label{diffn1}
\end{equation}
where $\vec{D}_i$ denotes the wave-function amplitudes on each slice in the spinful d-quasiparticle eigenbasis, and $A_i$ and $B_{i,i+1}$ are $2L^2 \times 2L^2$ matrices of the form
\begin{equation}
A_i = \left[\begin{array}{cc}t_{ii} & \Delta_{ii}\\ \Delta_{ii} & -t_{ii}\end{array}\right],
\end{equation}
\begin{equation}
B_{i,i+1} = \left[\begin{array}{cc}t_{i,i+1} & \Delta_{i,i+1}\\ \Delta_{i,i+1} & -t_{i,i+1}\end{array}\right],
\end{equation}
where the $t_{ij}$ and $\Delta_{ij}$'s are now matrices coupling slices 'i' and 'j' in the manner described by Eq.\ref{diffn1}. To obtain the transfer matrix, we rewrite Eq.\ref{diffn1} as 
\begin{equation}
\vec{D}_{i+1} = B^{-1}_{i,i+1}(E-A_i)\vec{D}_i - B^{-1}_{i,i+1}B_{i-1,i}\vec{D}_{i-1}.
\label{diffn2}
\end{equation}

For our model, the inter-slice coupling has the simple form
\begin{equation}
B_{i,i+1} = \frac{1}{\sqrt{2}}[I \otimes \sigma^x + I \otimes \sigma^z],
\label{d2}
\end{equation}
which satisfies the special property
\begin{equation}
B_{i,i+1} = B_{i,i+1}^{-1}.
\label{d3}
\end{equation}
Using Eq.\ref{diffn2}, we are now in a  position to define a transfer matrix as follows:
\begin{equation}
\left(\begin{array}{c} \vec{D}_{i+1} \\ \vec{D}_i \end{array}\right) = T_i\left(\begin{array}{c} \vec{D}_{i} \\ \vec{D}_{i-1} \end{array}\right),
\end{equation}
where $T_i$ has the relatively simple form
\begin{equation}
T_i = \left(\begin{array}{cc} B_{i,i+1}(E-A_i) & -I\\ I & 0 \end{array}\right),
\end{equation}
and the multiplicative property of Eq.\ref{Tmult}. With this transfer matrix at hand, which very closely resembles standard ones used for the Anderson model\cite{KMcK,Tero}, we can extract the localization length for different values of energy E and disorder W in a manner completely analogous to the numerical treatment of the Anderson model. As we are interested in the behaviour of states at the Fermi energy, we set the energy E, to zero.

The procedure for extracting the localization length is quite standard, and has been elaborated on in great depth in many works\cite{Sora,Tero}. But to briefly outline the method, one begins with an orthonormal basis of vectors $\hat{o_i}(0)$ in the space of the transfer matrix, representing the right-most states in Fig.\ref{quasi}. One then assumes that for a given disorder strength W, and width L, there exists a set of eigenmodes $\hat{w_i}(L,W)$ that describes typical eigenmodes for a quasi 1-dimensional system of N slices, where 'N' is large enough to represent the average behaviour of the random disorder, and that these modes decay or grow in magnitude as $\exp({\pm \gamma_i N})$ upon multiplication with the corresponding transfer matrix $T = \prod_{i=1}^N T_i$. One further assumes that when the set of basis vectors $\hat{o_i}(0)$ , each of which may be represented as a linear combination of the $\hat{w_i}(L,W)$,is multiplied by T, the resulting vectors $\vec{v_i}(1)$ (where the dependence on L and W is implicit), are each composed of the appropriate linear combination of modes $\hat{w_i}(L,W)$ now weighted by the corresponding growth factors $\exp({\pm \gamma_i N})$. The length 'N' is numerically restricted by the exponentially growing magnitude of the vectors $\vec{v}$.

 Since all vectors have now grown fastest along the fastest growing mode, say $\hat{w_1}(L,W)$, the magnitude and direction of any one of these modes, say $\vec{v_1}(1)$, are $\exp{\gamma_1 N}$ and along $\hat{w_1}(L,W)$ respectively. Projecting out the component $\vec{v_1}(1)$ from the next vector $\vec{v_2}(1)$ gives a resultant vector $\vec{o_2}(1)$ whose magnitude is roughly $\exp{\gamma_2 N}$. Thus, by such a 'Gram-Schmidt' orthogonalization procedure for the whole set of vectors $\vec{v_i}(1)$, one obtains a set of orthogonal vectors $\vec{o_i}(1)$ with associated Lyapunov exponents

\begin{equation}
\gamma_i(1) = \frac{\ln{|\vec{o_i}(1)|}}{N},
\label{Lyap}
\end{equation}
which give the characteristic inverse localization lengths associated with each mode. To reduce computational effort, we consider just the positive  Lyapunov exponents corresponding to exponentially growing states. As we are restricted in our length size N, to obtain a fair estimate of the typical $\gamma_i$'s, we repeat the procedure of transfer matrix multiplication, now using as our initial basis vectors the normalized set $\hat{o_i}(1)$ which more or less point along the ideal basis $\hat{w_i}(L,W)$. An average value $\frac{1}{M}\sum_{j=1}^M\gamma_i(j)$ obtained from M such iterations provides the desired estimate of the ideal $\gamma_i$'s. We associate the characteristic localization length $\lambda(L,W)$ with the slowest decaying mode, and thus with the inverse of the smallest positive Lyapunov exponent $\gamma_{min}(L,W)$. 

In the quasi 1-dimensional case, all modes are exponentially decaying or growing (corresponding to in-going or out-going states respectively) since even the slightest disorder is enough to localize states. But in the 3-dimensional limit, where the 2-dimensional cross-sectional area becomes large, we know that the modes ought to experience a transition from extended to localized behaviour as a function of disorder. To determine the critical disorder strength $W_c$ for this transition, the nature of the modes in 3-dimensions and the localization length for an infinite size system of given disorder strength, one can use a finite size scaling analysis of the quasi 1-dimensional system. The scaling function that we will use to do so is the dimensionless parameter
\begin{equation}
\Lambda(L,W) = \frac{\lambda(L,W)}{L}.
\label{dimp1}
\end{equation}
We now turn to the critical properties of the phase transition between the thermal insulator and thermal metal.

\section{Critical Properties}

\begin{figure}
\epsfxsize=3.5in
\centerline{\epsffile{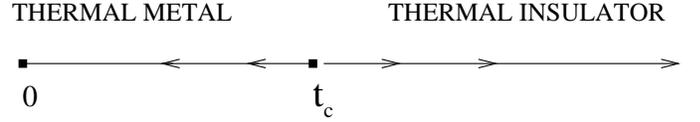}}
\vspace{0.15in}
\caption{Behaviour of coupling constant 't' with increasing system size,L.}
\vspace{0.15in}
\label{rgflow}
\end{figure}

As in disordered electronic systems, we have seen in Sec.\ref{FTF} that the thermal metal, the thermal insulator ,and the critical point seprating them may be characterized by their transport properties. The replica field theory of Eq.\ref{NLSM}, with its dimensionless coupling t, provides a powerful means of studying this transition. An analysis of the action in Eq.\ref{NLSM} shows that in $2+\epsilon$ dimensions, where $\epsilon=1$ for our system, an unstable fixed point $t_c$ describes the critical point between the thermal metal and thermal insulator. One can study the scaling behaviour of t with system size L(see Fig.\ref{rgflow}) by deriving a form for the scaling function $\beta(t)=\frac{dt}{d\ln{L}}$, which we present explicitly in the next section. Near $t_c$, for $t>t_c$, the coupling $t$ grows larger with L, and thus exhibits a stable thermal insulator, while for $t<t_c$, a smaller and smaller value of $t$ with increasing length scale signals a thermal metal.

\subsection{Localization Length Exponent}

The $\beta$ function for the $Sp(2n)$ sigma model of the action of Eq.\ref{NLSM} can be found in Ref.\cite{Hikami}, and it is given to cubic order('two' loop) in coupling t by
\begin{equation}
\beta(t;Sp(2n))= -\epsilon t + (2n+1) + \frac{1}{2}(2n+1)^2t^3 + {\cal O} (t^4),
\label{BS}
\end{equation}
where $\epsilon = d-2$, and n denotes the number of replicas. In the limit $n \rightarrow0$, at the critical point where the $\beta$ function vanishes, its derivative gives the inverse localization length exponent:
\begin{equation}
\frac{1}{\nu} = \epsilon + \frac{\epsilon^2}{2} + {\cal O} (\epsilon^3).
\label{nus} 
\end{equation}

In contrast, normal systems with time reversal symmetry and spin-rotational invariance may be identified with the $Sp(4n)/Sp(2n)\times Sp(2n)$ model of Ref.\cite{Hikami}, and its associated $\beta_n$ function has the form
\begin{equation}
\beta_n =  -\epsilon t +(4n+1)t^2 + (8 n^2 + 2 n)t^3 + {\cal O} (t^4).
\label{BN}
\end{equation}
The localization length exponent derived from Eq.\ref{BN} has the value
\begin{equation}
\frac{1}{\nu_n} = \epsilon + {\cal O} (\epsilon^3),
\label{nun} 
\end{equation}
quite different from the value of $\nu$ in Eq.\ref{nus}. The 
 numerical evidence to follow supports the field theoretic result that the value of the localization length exponent $\nu$ for the superconducting system with T and $SU(2)$ is considerably lower than the analogous exponent $\nu_n$ for normal systems.

\subsubsection{Numerical Treatment}
The standard numerical technique that we employ for extracting the localization length exponent $\nu$, shows that in three dimensions it takes on the value $1.15 \pm 0.15$. 

The finite-size scaling technique can be summarized as follows: scaling arguments require that the only relevant length scale in the system be the localization length $\xi(W)$ of the infinite sized system, and thus we have
\begin{equation}
\frac{\lambda(L,W)}{L} = \Lambda(L,W) = h(\frac{\xi(W)}{L}),
\label{dimp2}
\end{equation}
where h is a scaling function yet to be determined. Close to the critical point $W_c$, we have the localization length $\xi$ of the infinite system behaving as
\begin{equation}
\xi \sim |W-W_c|^{-\nu},
\label{critll}
\end{equation}
 which means that the argument '$x$' of $h(x)$ in Eq.\ref{dimp2} blows up at the critical point. However, $\Lambda$ is well-behaved and finite; this is only possible if we have
\begin{equation}
\lim_{x\to \infty} h(x) = const.,
\end{equation}
where the constant refers to independence with respect to L for large L. Thus, the critical value $\Lambda_c$ is common to all sufficiently large system sizes. Using Eq.\ref{critll}, we rewrite Eq.\ref{dimp2} as
\begin{equation}
\ln{\Lambda(L,W)} = f[L^{\frac{1}{\nu}}(W-W_c)].
\end{equation}
Linearizing the function 'f' about the critical fixed point $(W_c, \ln{\Lambda_c})$ yields
\begin{equation}
\ln{\Lambda(L,W)} = \ln{\Lambda_c} + A (W-W_c) L^{1/\nu}.
\label{dimp3}
\end{equation}

\begin{figure}
\epsfxsize=3.5in
\centerline{\epsffile{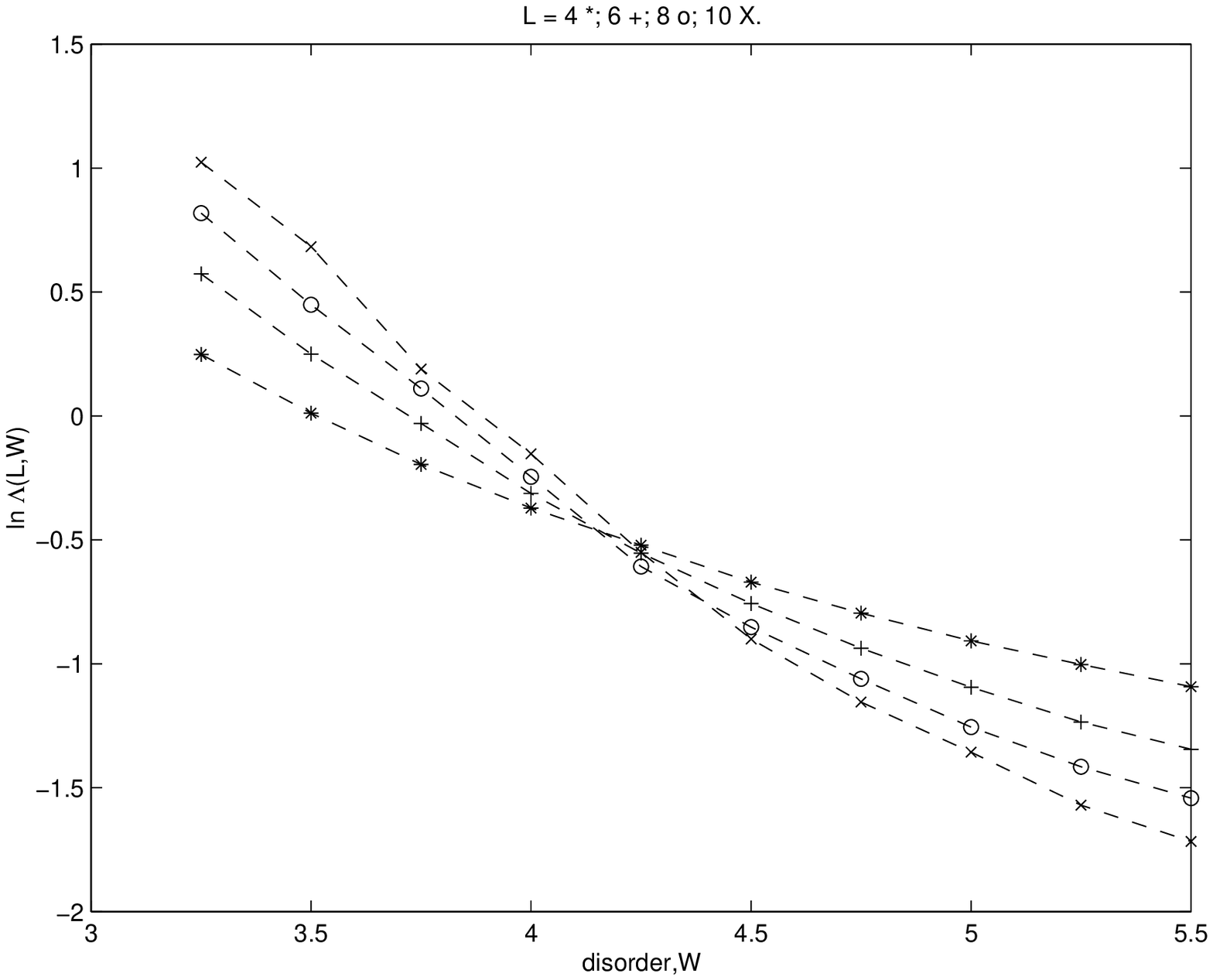}}
\vspace{0.15in}
\caption{The scaling function $\ln{\Lambda(L,W)}$ plotted as a function of disorder for different system sizes.}
\vspace{0.15in}
\label{lnlW}
\end{figure}

To procure the value of $\nu$, we  use an iterative procedure\cite{Tero} which is equivalent to the widely used procedure of Ref.\cite{KMcK} of performing a least square fit to obtain actual values of $\xi(W)$. We rewrite Eq.\ref{dimp3} as 
\begin{eqnarray*}
\ln{\Lambda} & = & AL^{1/\nu}W - (-\ln{\Lambda}_c + AL^{1/\nu}W_c)\\ &=&a(L)W - b(L,\Lambda_c,W_c),
\end{eqnarray*}
assume an initial value for the critical point $(W_c, \ln{\Lambda_c})$ from Fig.\ref{lnlW}, obtain the functions $a(L)$ and $b(L,\Lambda_c)$ by curve-fitting, extract $A$ and $\nu$, determine the value of the critical point thus obtained, and repeat the procedure till convergence is achieved (which happens rather quickly).If scaling is valid, we can collapse our data onto  curves of $\ln{\Lambda(W,L)}$ vs $L^{1/\nu}(W-W_c)$.

In our transfer matrix calculations, we choose systems whose cross-sectional areas have linear dimensions of $L =4,6,8,10$, and our transfer matrices have dimensions $4L^2\times4L^2$ with the given values of $L$. We choose the number of transfer matrices to be multiplied together by ensuring that the basis vectors do not grow upto a magnitude greater than $10^5$ upon being multiplied by the set of transfer matrices. We utilize a total of 2000 slices in the quasi 1-dimensional system for each value of L and W.

\begin{figure}
\epsfxsize=3.5in
\centerline{\epsffile{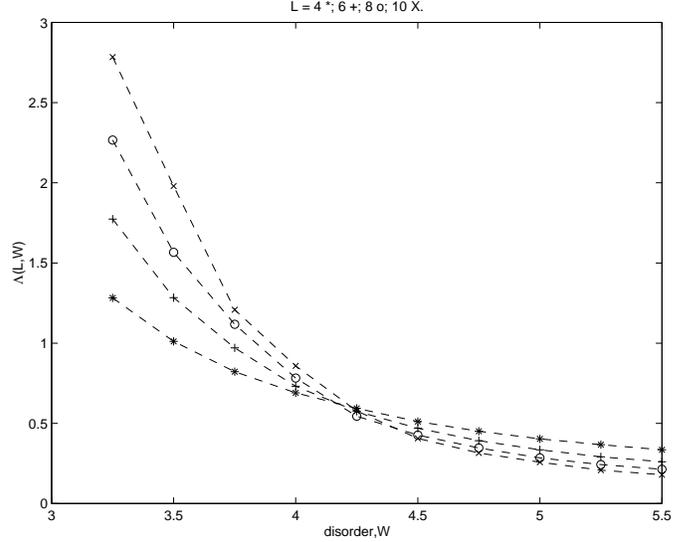}}
\vspace{0.15in}
\caption{The scaling function $\Lambda(L,W)$ plotted as a function of disorder for different system sizes.}
\vspace{0.15in}
\label{lamW}
\end{figure} 
Fig.\ref{lamW} shows the plots for $\Lambda(L,W)$ as a function of disorder for $L=4,6,8,10$. For fixed disorder W, an increasing $\Lambda(L,W)$ with increasing system size $L$ indicates the extended regime, while a decreasing $\Lambda(L,W)$ shows that the system is in the localized regime. In comparison to the Anderson model for normal systems, our simulations require a much smaller number of transfer matrices for relatively noise-free data, and we believe that this really is a consequence of the relatively low critical disorder strength.

Fig.\ref{lnlW} shows the data for the iterative procedure which gives the value for the localization length exponent

\begin{equation}
\nu = 1.15 \pm 0.15.
\end{equation}

\begin{figure}
\epsfxsize=3.5in
\centerline{\epsffile{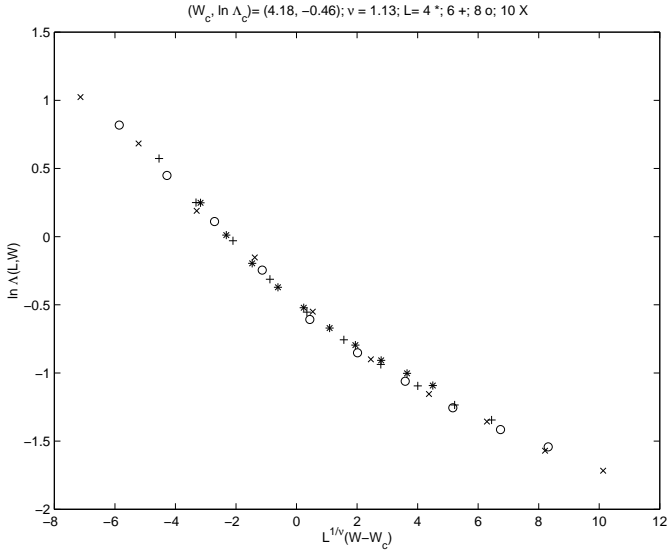}}
\vspace{0.15in}
\caption{Collapse of data for $\ln{\Lambda(L,W)}$ given the value $\nu = 1.13$ for the localization length exponent}
\vspace{0.15in}
\label{datcoll}
\end{figure} 
Finally, Fig.\ref{datcoll} indeed demonstrates clean data collapse close to the critical point. 

\subsubsection{Summary of Results}
 Both field theory and numerics concur with the fact that the localization length exponent in the superconducting systems is significantly lower than that of their normal partners, clearly indicating a new universality class. An $\epsilon$ expansion in $d=2+\epsilon$ dimensions of the action in Eq.\ref{NLSM} shows that the superconducting system has a localization length exponent $\nu$ of  $(\epsilon +\frac{\epsilon^2}{2} + {\cal O} (\epsilon^3))^{-1}$ in contrast to a $\nu_n$ of $(\epsilon +{\cal O} (\epsilon^3))^{-1}$ for normal systems. The field theory would thus predict $\nu = \frac{2}{3}$ versus $\nu_n=1$ in 3-dimensions. In comparison, one obtains the numerical estimate $\nu = 1.15 \pm 0.15$ versus $\nu_n = 1.54 \pm 0.08$\cite{KMcK} in 3-dimensions. 

We must remark that the system sizes and computing power utilized in our numerical studies were relatively low compared to the current cutting edge procedures. As a lot of work has gone into refining techniques with regard to normal systems(for e.g. Ref\cite{KMcK}), it is well worth employing them to study this novel phase transition and analogous ones in superconducting systems with other symmetries.

\subsection{Density of States}
\begin{figure}
\epsfxsize=3.5in
\centerline{\epsffile{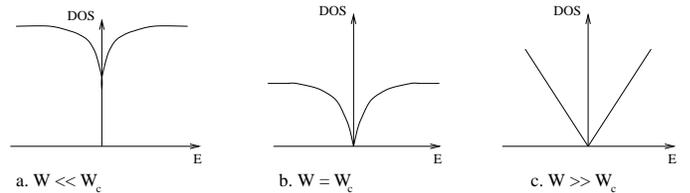}}
\vspace{0.15in}
\caption{Behaviour of DOS a) Well within the thermal metal b) At criticality c) Well within the thermal-insulator.}
\vspace{0.15in}
\label{dosbeh}
\end{figure}

The quasiparticle DOS in dirty superconducting systems  exhibits some of the most stunning differences from normal systems. In normal systems, both in the Anderson metal and the Anderson insulator, i.e., in the absence of interactions, the DOS remains a smooth continuous function across the Fermi energy. In contrast, in gapless superconductors that respect $SU(2)$, well within the thermal metal, quantum interference effects cause a singularity at the Fermi energy which manifests itself as a $\sqrt{E}$ cusp in 3-dimensional systems\cite{SVSF}. Deep in the thermal insulator, the density of states exhibits a power-law that vanishes at the Fermi energy with the form $\rho \sim |E|^{\alpha}$, where $\alpha=1$ for systems possessing time-reversal invariance\cite{TSMF2,SVSF}. About the critical point, the DOS once again shows power-law singularities, the details of which we discuss below. The curious form of the DOS (shown in Fig.\ref{dosbeh}) has profound impact on thermodynamic properties, and in particular, manifests itself in quantities such as specific heat and spin susceptibility.

\subsubsection{Discussion of Critical Behaviour}
The field theoretic action of Eq.\ref{NLSM} not only offers concrete predictions for the DOS, if one were to use Wegner's analogy with magnetic systems\cite{Wegner}, it provides an intuitive picture for the behaviour about criticality. To elaborate, the quasiparticle DOS at the Fermi energy, which also gives a measure of the magnetization, acts as the 'order parameter' of the field theory\cite{TSMF2}. It is given by 

\begin{equation}
\rho = \lim_{n\to 0}\frac{\rho_0}{4n}<Tr(U^{\dagger}+U)>,
\end{equation}
where $\rho_0$ is the bare DOS, and n the number of replicas. The field $\eta$, which has units of energy E, couples to the DOS in the action of Eq.\ref{NLSM}, and might be equated with the magnetic field in the magnetic analog.

With a little indulgence, one can go further with parallels between the field theory and the magnetic systems, as first suggested for normal systems\cite{Wegner}:
\bigskip

\begin{tabular}{|r|l|}
\hline
F.T. of $S_{N L \sigma M}$ & Magnetic Systems \\
\hline
Distance  from criticality & Reduced Temperature\\
$\Delta = \frac{W-W_C}{W_C}$ & t\\
\hline
DOS, $\rho$ & Magnetization, m\\
Energy, E & Magnetic Field, h\\
$\tilde{\chi} = \frac{d\rho}{dE}$ & Magnetic Susceptibility,$\chi$ \\
\hline
\end{tabular} \label{analog}

\bigskip
In normal systems, the analogy is clouded by the fact that the DOS is a continuous  function of energy and disorder respectively. One can reconcile with this if the DOS obeys a power-law form with exponent zero, and in fact, one can show this to be the case on field theoretic grounds\cite{Wegner}. But in  superconducting systems, as we shall see, the analogy goes through in quite a striking manner, with a whole slew of  nontrivial critical exponents:
\bigskip

\begin{tabular}{|c|}
\hline
$\rho(\Delta, E=0) \sim |\Delta|^{\beta}$\\
$\rho(\Delta=0, E) \sim |E|^{\frac{1}{\delta}}$ \\
$\tilde{\chi} \sim |\Delta|^{-\gamma}$ \\
$\xi \sim |\Delta|^{-\nu}$ \\
\hline
\end{tabular} \label{critexp}

\bigskip
where we have used the notation of the tables above,and $\xi$ is the localization length and $\nu$ the associated exponent described in the previous section. 

In order to derive expressions for $\beta$, $\delta$ and $\nu$, we start with the 'free energy density' f, obtained from the action of Eq.\ref{NLSM} :

\begin{equation}
f  =  \frac{1}{L^d}\lim_{n \to 0}\frac{ln{\cal Z}_n}{n},
\end{equation}
\begin{equation}
{\cal Z}_n  = \overline{Z^n} = \int d \tilde{U} e^{-S_{N L \sigma M}},
\end{equation}
 where $d \tilde{U}$,the integral volume element,  takes into account  the symplectic group structure of the matrices in the action $S_{N L \sigma M}$, n denotes the number of replicas, $Z$ denotes the partition function of single system, and the overbar above $Z^n$ refers to the average over disorder. Near criticality, $f_s$, the singular part of the free energy density, is expected to scale as follows:

\begin{equation}
f_s(\Delta, E) = \xi^{-d}\tilde{f}(\xi^yE),
\label{fsing}
\end{equation}

where 'y' describes the scaling form of E, and Eq.\ref{critll} gives the behaviour of the correlation length $\xi \sim |\Delta|^{-\nu}$. 
Differentiating the free energy with respect to E results in the following form for the DOS:
\begin{equation}
\rho(\Delta,E) = \xi^{-d+y}F^{\pm}(\xi^y E),
\label{DOS}
\end{equation}
where $F^+$ corresponds to behaviour for $\Delta>0$, and $F^-$ for $\Delta<0$. To obtain  $\beta$, we set $E=0$, and compare the form of the resulting order parameter $\rho(\Delta, E=0)$ in the above table, yielding

\begin{equation}
\beta = \nu(d-y).
\label{bet}
\end{equation}

Here, we require $F^-(0)$ to be finite, and $F^+(0)=0$. 

To extract $\delta$, we impose the physical constraint that $\rho$ be well-behaved and finite at criticality. This requires that $\rho(\Delta \to 0,E)$ be independant of the diverging correlation length, and thus yields
\begin{equation}
\frac{1}{\delta} = \frac{d}{y}-1.
\label{del}
\end{equation}
Finally, taking a derivative of $\rho$ in Eq.\ref{DOS} with respect to E gives us the following expression for $\gamma$: 
\begin{equation}
\gamma = \nu(2y-d).
\label{gam}
\end{equation}

\subsubsection{Results}
To obtain estimates of critical exponents from field theoretic results in $2+\epsilon$ dimensions, we use the value $(\epsilon + \frac{\epsilon^2}{2})^{-1} + {\cal O} (\epsilon^3)$ obtained for $\nu$ in the previous section, and the value $\frac{1}{\delta} = \frac{\epsilon}{4} + {\cal O}(\epsilon^3)$ from Ref.\cite{TSMF2}. Eq.\ref{bet}-\ref{gam} then enable us to determine the critical exponents $\beta$ and $\gamma$ via the relationship $\frac{4d}{\epsilon + 4}$.
 Specifically, in the case of 3-dimensions, substituting the value $\epsilon = 1$, we obtain the rough estimates $y=12/5$, $\delta = 4$, $\nu = 2/3$, $\beta = 2/5$ and $\gamma = 6/5$.

Shifting our focus to numerical results, the  method of exact diagonalization reveals that the superconducting system at hand does indeed show the novel singular behaviour in the DOS at the Fermi energy. In the data shown below, we have once more modelled the superconducting Hamiltonian after Eq.\ref{HdL} using periodic boundary conditions. Systems of linear dimension 'L' have required matrices of dimension $2 L^3 X 2 L^3$, and we have explored system sizes with linear dimensions $L=4,6,8$.

\begin{figure}
\epsfxsize=3.5in
\centerline{\epsffile{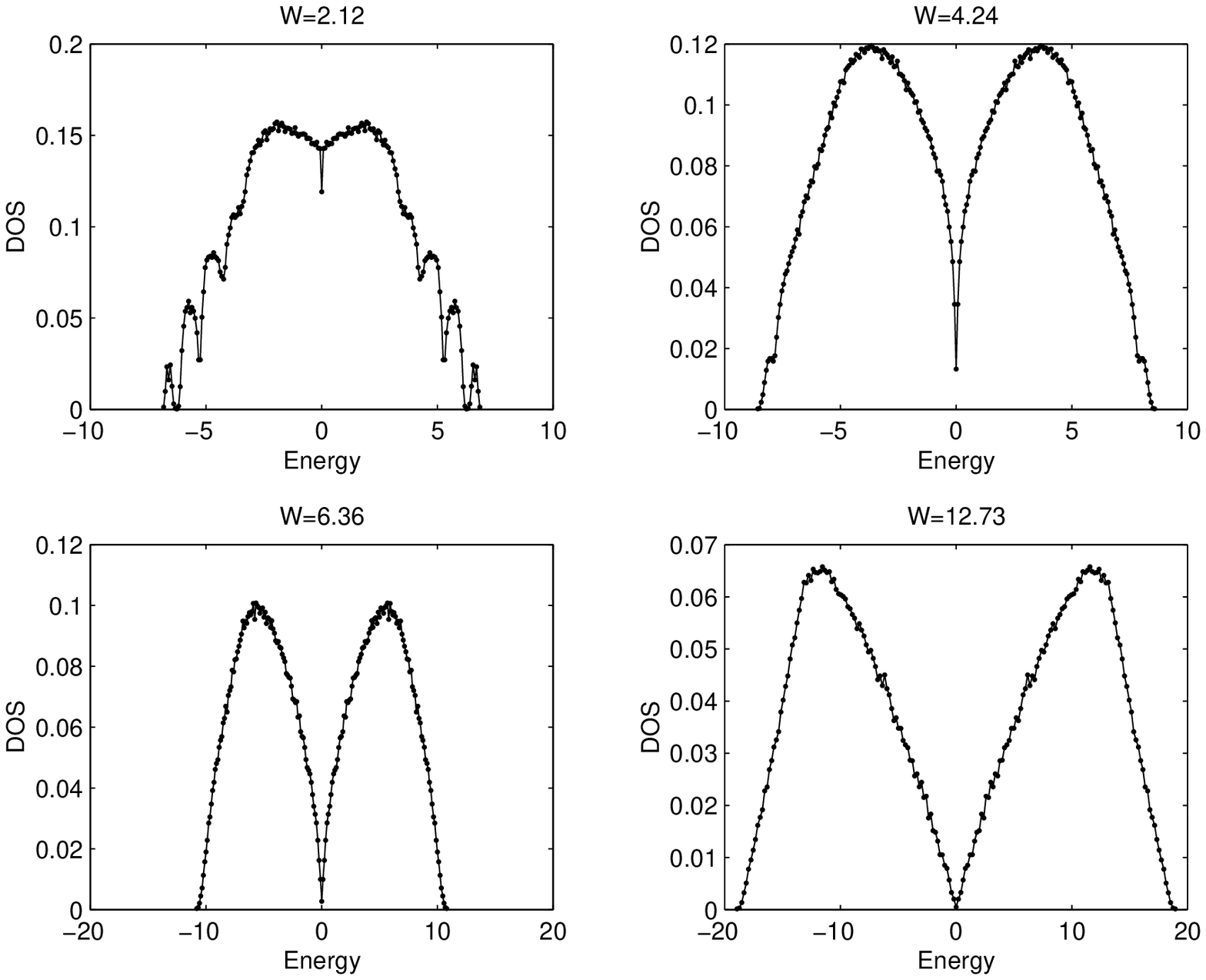}}
\vspace{0.15in}
\caption{Progression of DOS as a function of energy for various disorder strengths.}
\vspace{0.15in}
\label{dosprog}
\end{figure}

Fig.\ref{dosprog} shows the progression of the behaviour of the DOS with increasing disorder strength. As seen in the last panel of Fig.\ref{dosprog}, the DOS shows a power-law behaviour of the form $\rho \sim |E|$ about the Fermi energy , $E_F$, consistent with expectations for the thermal insulator.

\begin{figure}
\epsfxsize=3.5in
\centerline{\epsffile{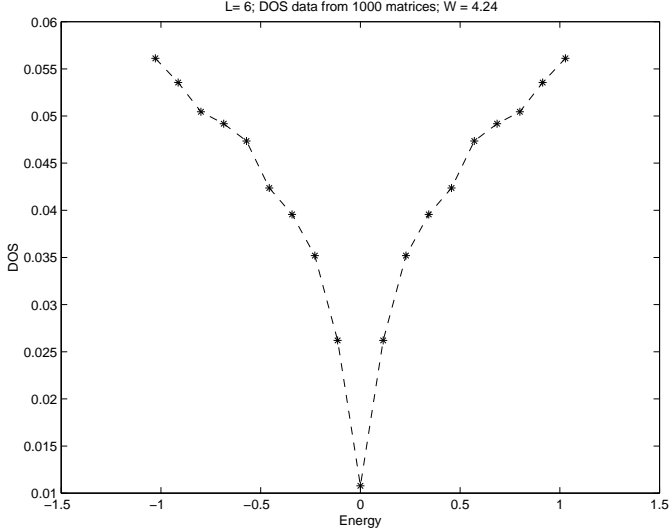}}
\vspace{0.15in}
\caption{Zoom of DOS about $E=0$ for $W=4.24$.}
\vspace{0.15in}
\label{doszoom}
\end{figure} 

Fig.\ref{doszoom} shows a zoom of the DOS about $E_F$ for disorder close to the critical strength $W_C$, for which we have an estimate from the localization length study of the previous section. One can easily discern that the DOS plummets down quite markedly, and does indeed exhibit singular power-law behaviour.

\begin{figure}
\epsfxsize=3.5in
\centerline{\epsffile{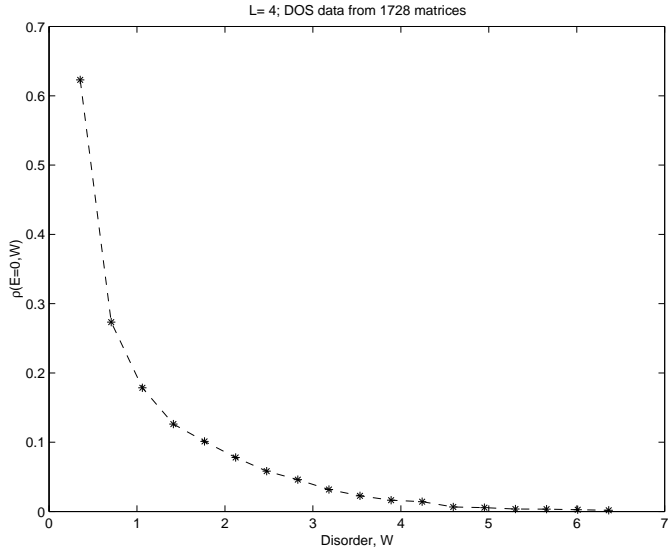}}
\vspace{0.15in}
\caption{DOS at E=0 as a function of disorder strength.}
\vspace{0.15in}
\label{dosor}
\end{figure} 

A plot of the DOS at the Fermi energy(Fig.\ref{dosor}) shows that even relatively small system sizes provide numerical confirmation of the fact that $\rho(E=0)$ acts as the order parameter for the field theory of Eq.\ref{NLSM}; the DOS at $E_F$ is finite for low disorder, and it slowly drops to zero beyond a critical disorder strength. As discussed in the previous section, one would in fact expect the DOS for an infinite sized system to behave as $\rho(\Delta, E=0) \sim |\Delta|^{\beta}$, where $\Delta$ is the distance from criticality within the thermal metal. Scaling arguments for extracting $\beta$ require
\begin{equation}
\rho_L(\Delta, E=0) = |\Delta|^{\beta}Y(L \Delta^{\nu}),\ \Delta < 0,
\label{doscl1}
\end{equation}
where $\rho_L$ is the DOS associated with a system of linear dimension L, and $Y$ is a scaling function. One can rewrite the above equation in a form more conducive to numerics as follows:
\begin{equation}
\rho_L(\Delta,E=0) = L^{-\frac{\beta}{\nu}}\tilde{Y}(\Delta L^{\frac{1}{\nu}}),
\label{doscl2}
\end{equation}
where $\tilde{Y}$ is yet another scaling function with limiting behaviour $\tilde{Y}(x \to \infty) = |x|^{\beta}$, reproducing the required dependence of $\rho(\Delta, E=0)$ on $\Delta$ for infinite system size. 

\begin{figure}
\epsfxsize=3.5in
\centerline{\epsffile{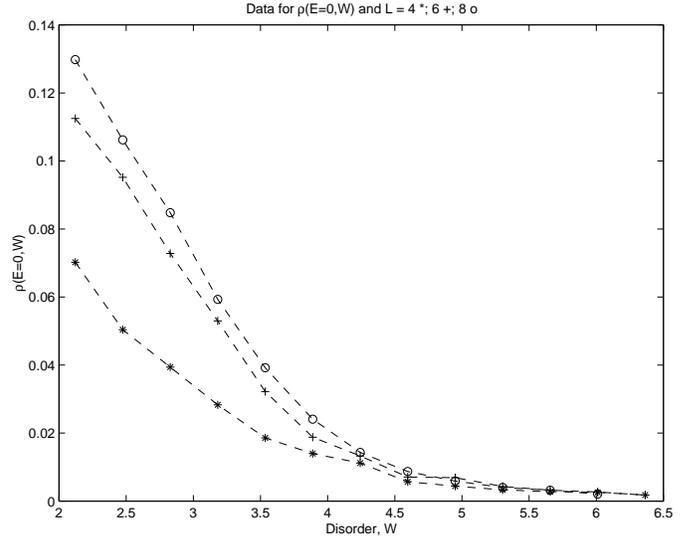}}
\vspace{0.15in}
\caption{DOS at E=0 for various disorder strengths W, and system sizes L}
\vspace{0.15in}
\label{dosL}
\end{figure} 

Fig.\ref{dosL} exhibits the plots of $\rho_L$ as a function of disorder for different system sizes L, and we make use of this data to procure the value of $\beta$ in Eq.\ref{doscl2}. To extract $\beta$, we perform a fit taking $W_c$, $\nu$ and $\beta$ as variable parameters. Exploiting the universal nature of the function $\tilde{Y}$ in Eq.\ref{doscl2}, we find the appropriate values of $\rho$ obtained by linear interpolation for a given set of system sizes and fixed argument in $\tilde{Y}$, and plot these on a log-log scale versus system size; the slope for a linear fit of such a set of points determines $\beta$. The actual value of  $\beta$ is obtained by performing the above procedure for different values of the argument of $\tilde{Y}$ and taking the average of the $\beta$'s thus obtained.

\begin{figure}
\epsfxsize=3.5in
\centerline{\epsffile{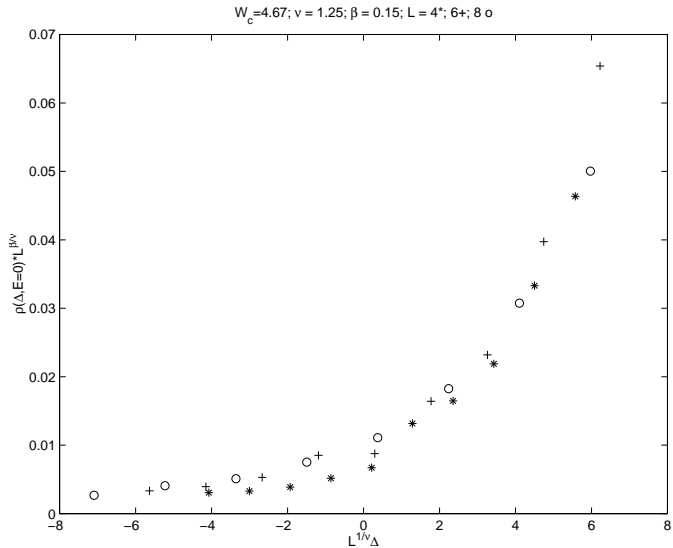}}
\vspace{0.15in}
\caption{Data collapse of $\rho(E=0,W)$ given the form Eq.\ref{doscl2}, and the critical exponent $\beta=0.15$.}
\vspace{0.15in}
\label{dscoll}
\end{figure} 

The set of values $\beta = 0.15$, $W_c=4.67$ and $\nu=1.25$ result in the data collapse shown in Fig.\ref{dscoll}. In comparison, as remarked at the beginning of this section, the field theoretic result predicts that $\beta=0.4$. Once more, as in the case of the localization length exponent, we comment on numerical accuracy; other simulations using exact diagonalization, for instance, those catering to specific physical situations\cite{MacD}, have used larger sytems sizes and number of realizations which would be well-worth employed here. However, the above numerics conveys quite clearly that the DOS at criticality exhibits a power law suppression about the Fermi energy, and that $\rho(E=0,W)$ acts as an 'order parameter' with a non-trivial exponent $\beta$ in surprising contrast to normal systems which have $\rho(E ,W_c)$ smooth about $E_F$, and a vanishing exponent $\beta$.

\section{Other Systems}
We have studied the thermal metal-thermal insulator transition in superconducting systems with $SU(2)$ and $T$, and discussed the dirty d-wave superconductor as a possible physical realization. Superconducting systems with other symmetries too promise such a transition.

The properties and phases of superconducting systems with spin-rotational invariance, but no time reversal invariance are rather similar to our case which preserves both symmetries. The thermal metal and the thermal insulator are both stable only in 3-dimensions, and the transition cannot occur in lower dimensions where quasi-particle excitations are always found to be localized at large enough length scales. Due to the absence of time-reversal symmetry, the Hamiltonian of Eq.\ref{HL} now has imaginary couplings. As described in detail in previous work\cite{SVSF}, in this symmetry category, the pinned vortex state of a type II s-wave superconductor appears to be a fine candidate for exhibiting the thermal metal-thermal insulator transition. Low energy quasiparticles bound to the core of the vortices can tunnel from one vortex to another, and as the magnetic field is increased, the density and tunneling strength also increase. It is conjectured that there could exist a critical magnetic field $H_{c4}$, within the vortex phase at which the low energy states can permeate through the medium to form extended states. Field theoretic methods have analyzed the properties of these systems, and in particular, have shown that in parallel to the case with T, the density of states at the critical point has the power-law behaviour $\rho \sim |E|^{\epsilon/2}$, where $E$ is the energy and $\epsilon = d-2$.

The presence of spin-orbit scattering or of triplet pairing introduces new ingredients. It breaks spin-rotational symmetry, and as in the case of normal systems, field theoretic arguments predict the presence of a delocalization-localization transition not only in 3-dimensions but also in 2-dimensions\cite{TSMF4}.In Ref.\cite{TSMF4}, T.Senthil et.al. cast their Hamiltonian in terms of Majorana fermions, and their formulation, among other things, is highly conducive to numerical work. They discuss possibilities for experiment, and mention heavy fermion systems where spin-orbit scattering is prominent, as a physical realization.

Finally, the tantalizing prospect of a Hall effect in superconducting systems has been explored in systems with other symmetries as well,i.e., those with SU(2)\cite{TSMF3}. Ref\cite{TSMF3} shows that a superconductor with unconventional $d_{x^2-y^2}+id_{xy}$ pairing symmetry is capable of exhibiting a phase with non-vanishing spin and thermal Hall conductances. Indeed, as in normal systems, sophisticated methods such as employing the network model and supersymmetric spin chains, have shed light on these systems\cite{Ilya}.

\section{Experiment}
The transport properties of normal systems have been probed in great depth, and now we see that superconducting systems could potentially offer an equally rich range of experiments in the thermal metal-thermal insulator transition. In the previous section, we have mentioned a variety of experimental candidates for study, such as the dirty d-wave, the type II s-wave, heavy fermion and other superconducting systems. These systems must share the feature of gapless superconductivity; one requires states at the Fermi energy since it is these states that determine transport properties, and distinguish the thermal insulator from the thermal metal. Associated with each system, a tunable parameter such as disorder or magnetic field ought to be able to access the phases. As was previously discussed in detail\cite{SVSF}, the type II s-wave superconductor in the vortex phase offers promise as a likely candidate for observing the novel transition since in principle, one need only tune the magnetic field, and generally the vortex phase exists over a large range of field.

As seen earlier, in contrast to normal systems, the density of states shows singular behaviour about the Fermi energy for both phases and at the critical point. We saw that for systems with T and SU(2), it obeys a power-law behaviour of the form $\rho(E) \sim |E|^{\alpha}$, where $\alpha = \frac{1}{2}$ well within the thermal metal, $\alpha = 1$ deep in the thermal insulator, and field theory predicts $\alpha = \frac{\epsilon}{4}$ at the transition, with $\epsilon=1$ for 3-dimensional systems. This singular behaviour ought to be reflected in thermodynamic quantities such as specific heat and spin susceptibility, and in tunneling experiments. In particular, the temperature dependence of the thermodynamic quantities would have a form $C \sim T^{1+\alpha}$ for the specific heat, and $\chi \sim T^{\alpha}$ for the spin susceptibility. Recent experiments of cuprate superconductors have observed a suppression of the specific heat close to the Fermi energy\cite{JYL}. However, measuring the differing behaviours to detemine the phase might prove tricky, especially since we have neglected various effects such as interactions which could  come into play.

For systems that preserve spin-rotational invariance, quasiparticle excitations about the superconducting ground state not only conserve energy, but also spin; the spin-conductance $\sigma_s$ could be employed to determine whether the system inhabits the thermal metal or thermal insulator. In the thermal metallic phase, a magnetic field gradient would cause the spinful quasiparticle to diffuse across the sample, while in the thermal insulator, they would be unable to conduct spin. A variety of refined spin-injection techniques have been developed in semi-conductors to measure spin dynamics (see for e.g. Ref.\cite{Awsch}), but by no means would it be a simple task to cater these experiments to superconductors.

We believe that by far, thermal conductivity measurements would offer most promise in probing the thermal metal-thermal insulator transition. For all superconducting systems with their differing symmetries, the thermal conductivity $\kappa$, distinguishes the two phases in that the ratio $\kappa/T$ tends to a finite constant in the thermal metal, and to zero as $T\rightarrow 0$ in the thermal insulator. Along the lines of previous experiments\cite{Taill}, it would be extremely interesting to observe the transition by applying a small thermal gradient across a superconducting sample and measuring $\kappa/T$ as a function of a tunable parameter.

We thank T.Senthil and I.A. Gruzberg for many an illuminating conversation, and D. Whysong for indispensible advice on numerical work. This research was supported by NSF Grants DMR-97-04005, DMR95-28578 and PHY94-07194.

\end{multicols}

\begin{references}

\bibitem{PLTVR} P.A.Lee and T.V.Ramakrishnan, Rev. Mod. Phys. {\bf 57}, 287 (1985).

\bibitem{Bodo} B. Huckstein, Rev. Mod. Phys. {\bf 67}, 357 (1995), and references therein.

\bibitem{Bund} R. Bundschuh, C.Cassanello, D. Serban and M.R. Zirnbauer, Phys. Rev. B {\bf 59} 4382 (1999); R. Bundschuh, C. Cassanello, D. Serban and M.R. Zirnbauer, Nucl. Phys. B {\bf 532}, 689 (1998). 

\bibitem{TSMF1} T.Senthil, M.P.A. Fisher, L.Balents and C.Nayak, Phys. Rev. Lett. {\bf 81}, 4704 (1998).

\bibitem{AZ} A. Altland and M.R. Zirnbauer, Phys. Rev. B {\bf 55}, 1142 (1997).

\bibitem{Opp} R. Oppermann, Physica A {\bf 167}, 301 (1990).

\bibitem{Ilya} V. Kagalovsky, B. Horovitz, Y. Avishai and J.T. Chalker, Phys. Rev. Lett. {\bf 82}, 3516 (1999); I.A. Gruzberg, A.W.W. Ludwig and N. Read, Phys. Rev. Lett. {\bf 82} 4524 (1999).  

\bibitem{TSMF2} T. Senthil and M.P.A. Fisher, Phys. Rev. B {\bf 60}, 6893 (1999)

\bibitem{TSMF3} T. Senthil, J.B. Marston and M.P.A. Fisher, Phys. Rev. B {\bf 60} ,4245 (1999). 

\bibitem{SVSF} S. Vishveshwara, T. Senthil, and M.P.A. Fisher, Phys. Rev. B {\bf 61}, 6966 (2000). 

\bibitem{TSMF4} T. Senthil and M. P. A. Fisher, cond-mat/9906290. 

\bibitem{Gork} L. P. Gor'kov and P. A. Kalugin, JETP Lett., {\bf 41}, 253 (1985)

\bibitem{KMcK} A. MacKinnon and B. Kramer, Z. Physik B {\bf 53}, 1 (1983); B. Kramer and A. MacKinnon, Rep. Prog. Phys. {\bf 56}, 1469 (1993); A. MacKinnon, J. Phys. {\bf 6}, 2511 (1994).


\bibitem{Sora} S. Cho, Ph.D. Thesis, UC Santa Barbara, unpublished.


\bibitem{Been} C.W.J. Beenakker, Rev. Mod. Phys. {\bf 69}, 731 (1997).

\bibitem{Caselle} M.Caselle, cond-mat/9610017. 


\bibitem{Tero} T. Heikkila, Helsinki Univ. of Tech., unpublished. 

\bibitem{Hikami} S. Hikami, Phys. Lett. {\bf 98 B}, 208 (1981).

\bibitem{Wegner} F. J. Wegner, Z. Physik.B {\bf 35}, 207 (1979); A. J. McKane and M. Stone, Ann. Phys. {\bf 131}, 36 (1981); F. J. Wegner, Z. Physik B {\bf 25}, 327 (1976)

\bibitem{MacD} W. A. Atkinson, P. J. Hirschfeld, A. H. MacDonald, cond-mat/0002333. 


\bibitem{JYL} C. F. Chang, J. Y. Lin and H. D. Yand, cond-mat/0003022 (2000).


\bibitem{Awsch} D. D. Awschalom, N. Samarth, J. Magn. Magn. Mater {\bf 200}, 130 (1999).


\bibitem{Taill} See for eg., N. P. Ong, K. Krishana, Y.Zhang and  Z. A. Xu, cond-mat/9904160 and references therein. 





\end{references}
\end{document}